\documentclass[aps,floats,preprint]{revtex4}
\usepackage{amsmath,amssymb,bm}
\usepackage[dvips]{color}
\usepackage{graphicx}

\newcommand{\al}{\alpha}
\newcommand{\ga}{\gamma}

\newcommand{\De}{\Delta}
\newcommand{\ep}{\varepsilon}

\newcommand{\eps}{\epsilon}

\newcommand{\la}{\lambda}

\newcommand{\si}{\sigma}

\renewcommand{\th}{\theta}   

\newcommand{\om}{\omega}

\newcommand{\ka}{\kappa}

\newcommand{\p}{\partial}
\newcommand{\<}{\langle} 
\renewcommand{\>}{\rangle} 
\newcommand{\txt}{\textstyle}
\newcommand{\dsp}{\displaystyle}
\newcommand\eqn[1]{(\ref{#1})}      
\newcommand{\beq}{\begin{equation}}
\newcommand{\eeq}{\end{equation}}
\newcommand{\be}{\begin{equation}}
\newcommand{\ee}{\end{equation}}
\newcommand{\ba}{\begin{array}}
\newcommand{\ea}{\end{array}}
\newcommand{\bc}{\begin{center}}
\newcommand{\ec}{\end{center}}
\newcommand{\bea}{\begin{eqnarray}}
\newcommand{\eea}{\end{eqnarray}}
\newcommand{\bi}{\begin{itemize}}  
\newcommand{\ei}{\end{itemize}}
\newcommand{\ben}{\begin{enumerate}} 
\newcommand{\een}{\end{enumerate}}

\newcommand{\half} {{\txt \frac{1}{2}}}
\newcommand{\third}{{\txt \frac{1}{3}}}

\newcommand{\twothirds}{{\txt \frac{2}{3}}}

\newcommand{\ie}{{\it i.e.}\,}
\newcommand{\eg}{{\it e.g.}\,}

\newenvironment{tightlist}[1]{ 
\begin{list}{#1}{
  \usecounter{enumi}
  \setlength{\topsep}{0ex} 
  \setlength{\itemsep}{-\parsep} 
  \settowidth{\labelwidth}{#1} 
  \setlength{\labelsep}{0.5em}        
  \setlength{\leftmargin}{\labelwidth}
  \addtolength{\leftmargin}{\labelsep}
 }}{\end{list}}

\newcommand{\fm}{{\rm fm}}
\newcommand{\km}{{\rm km}}
\newcommand{\eV}{{\rm eV}}

\newcommand{\MeV}{{\rm MeV}}

\newcommand{\K}{{\rm K}}
\newcommand{\G}{{\rm G}}

\newcommand{\Qt}{{\tilde Q}}
\newcommand{\X}{X}
\newcommand{\eX}{{e^{(\X)}}}
\newcommand{\eQt}{{e^{(\Qt)}}}
\newcommand{\alQt}{{\tilde\alpha}}
\newcommand{\mixang}{{\varphi}}
\newcommand{\kaX}{\ka_{\!X}}

\newcommand{\vecp}{\bm p}

\newcommand{\vecu}{\bm u}
\newcommand{\vecf}{\bm f}
\newcommand{\vecnabla}{\bm \nabla}

\newcommand{\vecH}{\bm H}
\newcommand{\vecB}{\bm B}
\newcommand{\vecA}{\bm A}
\newcommand{\vecj}{\bm j}

\newcommand{\veck}{\bm k}
\newcommand{\vecv}{\bm v}

\newcommand{\kperp}{k_{\perp}}
\newcommand{\pperp}{p_{\perp}}

\begin{document}

\title{
Color-magnetic flux tubes in quark matter cores of neutron stars}
\author{Mark G. Alford}
\email{alford@wuphys.wustl.edu}
\affiliation{Physics Department, Washington University,
St.~Louis, MO~63130-4899, USA}
\author{Armen Sedrakian}
\email{sedrakian@th.physik.uni-frankfurt.de}
\affiliation{Institute for Theoretical Physics, 
J. W. Goethe-University, D-60438 Frankfurt am Main, Germany}

\begin{abstract}
We argue that if color-superconducting 
quark matter exists in the core of a neutron star, it may
contain a high density of flux tubes, carrying flux that
is mostly color-magnetic, with a small admixture of ordinary 
magnetic flux. We focus on the two-flavor color-superconducting 
(``2SC'') phase, and assume that the flux tubes
are energetically stable, although this has not yet been demonstrated.
The density of flux tubes depends on the nature of the transition
to the color-superconducting phase, and could be within
an order of magnitude of the density of magnetic flux tubes
that would be found if the core were superconducting nuclear matter.
We calculate the cross-section for Aharonov-Bohm scattering of
gapless fermions off the flux tubes, and the associated collision
time and frictional force on a moving flux tube. We discuss
the other forces on the flux tube, and find that
if we take in to account only the forces that arise within
the 2SC core region then the timescale for expulsion of the color flux
tubes from the 2SC core is of order $10^{10}$ years. 

\end{abstract}

\date{10 Feb 2010} 


\pacs{97.60.Jd,12.38.Mh,47.32.C-,03.65.Ta,67.10.Jn}

\maketitle
\section{Introduction}
\label{sec:intro}

It has long been conjectured that neutron stars might contain
cores of quark matter, and one of the challenges facing
nuclear astrophysics is to find signatures by which the
presence of such matter could be inferred from observations
of the behavior of neutron stars. This requires us to
develop a good understanding of the differences between the
properties of nuclear matter and quark matter, taking in to
account the effects of magnetic fields, which are known to be 
present in neutron stars. In this paper we study quark matter
in magnetic fields $B\lesssim 10^{14}$\,Gauss, which are
astrophysically plausible and
high enough to affect transport (see for example \cite{Huang:2009ue}) 
but not so large as to modify the phase structure of the material 
\cite{Ferrer:2006vw,Fukushima:2007fc,Menezes:2008qt}.

Nuclear matter at high densities and low temperatures
is expected to be a type-II
electrical superconductor, with the magnetic field distributed
in an Abrikosov lattice of flux tubes~\cite{Baym:1969}.
In this paper we
investigate the possibility that quark matter in the two-flavor color
superconducting phase (``2SC'') \cite{Alford:2007xm} could be a type-II
superconductor with respect to the color gauge fields
\cite{Iida:2002ev,Giannakis:2003am}, with color flux tubes 
that scatter electrons, muons, and ungapped quarks
via the Aharonov-Bohm effect. 
These tubes are not topologically stable, and their energetic stability
has not yet been determined; in this paper we investigate the role they
might play in transport, and their expulsion time, if they turn out
to be stable or to have a lifetime that is sufficiently long.
As we explain below, the tubes carry flux that is mostly color-magnetic (hence they
can reasonably be called ``color-magnetic flux tubes'') with a
small admixture of ordinary magnetic flux.
We will argue that the density of color-magnetic flux tubes 
could be high, perhaps only about an order of
magnitude less than that of ordinary flux tubes in
superconducting nuclear matter.
Color magnetic flux tubes
may appear in other color superconducting phases, such as the 
color-flavor-locked (CFL) phase, but the CFL phase has no gapless
charged excitations, so in this paper we focus on the 2SC phase.
We calculate the Aharonov-Bohm interaction between the flux tubes
and unpaired quarks or electrons/muons.
We calculate the associated damping time and the forces on the
flux tubes. We defer the calculation of other contributions to relaxation
and transport in the 2SC phase,
such as scattering of the unpaired quarks and electrons off each other,
to future work.

The behavior of quark matter phases in magnetic fields is complicated 
by the intertwined breaking of the strong interaction
$SU(3)$ ``color'' gauge symmetry and
the electromagnetic $U(1)_Q$ gauge symmetry. In the 2SC phase, a condensate
of Cooper pairs of up ($u$) and down ($d$) quarks leads to the
gauge symmetry breaking pattern $SU(3)\otimes U(1)_Q \to 
SU(2)_{rg} \otimes U(1)_\Qt$ \cite{Alford:1997zt,Alford:1999pb}.
The unbroken $SU(2)_{rg}$ symmetry ensures confinement of 
particles that carry net red or green color, with a confinement
scale around 10\,MeV \cite{Rischke:2000cn}.
The unbroken $U(1)_\Qt$ gauge symmetry is a linear combination of
the original electromagnetic and color symmetries, called ``rotated
electromagnetism''. The
associated gauge field, the ``$\Qt$ photon'',
is a combination of the original photon and one of the gluons.
It is massless and propagates freely in 2SC quark matter.
The orthogonal combination $\X$ is a broken gauge generator, and
the associated magnetic field has a finite penetration depth.

The situation is closely analogous to the Higgs mechanism in the
standard model, where one linear combination of the hypercharge
and $W_3$ gauge bosons remains massless (the photon), while the
orthogonal combination becomes massive (the $Z^0$).
The $\X$ flux tubes are therefore analogous to ``$Z$-strings''
\cite{Vachaspati:1992fi} which have been found to be stable
only in a small region of the standard model parameter space
\cite{James:1992wb}, although the stable region may be enlarged when
bound states are taken in to account \cite{Vachaspati:1992mk}.
There are differences between the
2SC phase of QCD and the Higgs phase of the standard model:
the gluon mass is proportional to the quark chemical potential, not
the superconducting order parameter \cite{Alford:2007xm}; 
the non-Abelian gauge group
is $SU(3)$ rather than $SU(2)$ and is only partly broken, 
leaving an unbroken confining $SU(2)$ as well as an unbroken
$U(1)$ in the low temperature phase.
This means that a separate stability calculation will be needed
for the 2SC case.

Because electromagnetism is much more weakly coupled than the strong
interaction, the massless $\Qt$ gauge field is almost identical
to the photon, with a small admixture of a color gauge boson.
Conversely, the broken $\X$ gauge field is almost identical
to one of the gluons, with a small admixture of the photon 
\cite{Alford:1999pb}. Thus the $\X$ flux tubes can be described
as ``color-magnetic flux tubes''. However, because they contain a
small admixture of ordinary magnetic flux, they interact with
electrons/muons as well as with unpaired (blue) quarks.
In summary, the 2SC phase is not a superfluid, but it is
a superconductor with respect to
the $X$ gauge fields, and a conductor with respect to the $\Qt$
gauge fields, with current mainly being carried by the gapless 
electrons and blue quarks (one of which is neutral, the other has charge +1).
Strange quarks and muons, if present, will have a lower Fermi momentum
because of their higher mass, and hence less phase space near their Fermi 
surface. Thus their contribution to the processes discussed in this paper
will be subleading, and we ignore it.

The picture given above is valid below the critical temperature
for 2SC pairing and above an unknown critical temperature $T_{1SC}$
at which there will be a transition to a phase in which there
is self-pairing of the blue up and down quarks.
Such pairing would break the $U(1)_\Qt$ symmetry, so there could be
both $\Qt$ and $X$ flux tubes. Models of the strong interaction
between quarks do not give us much idea of the value of $T_{1SC}$.
They agree that, because the strong attraction is much weaker
in the single-color channel, $T_{1SC}$ will be many orders of
magnitude lower than the critical temperature for 2SC pairing,
perhaps as low as 1\,eV ($10^4$\,K)
\cite{Alford:1997zt,Schafer:2000tw,Alford:2002rz}. In this paper
we will be concerned with temepratures above $T_{1SC}$, where
the $\Qt$ gauge symmetry remains unbroken.

Depending on the ratio of the $\X$-flux penetration depth to the
coherence length of the condensate, the 2SC phase may be
type-I or type-II with respect to the $\X$ magnetic field
\cite{Iida:2002ev}.
In this paper we will be concerned with the possibility of
type-II behavior, and the presence of flux tubes containing 
$\X$-flux in the 2SC quark matter core
of a compact star. Even if the average magnetic field strength in the core
is below the lower critical field, such flux tubes may
end up ``frozen in'' if the quark matter
had cooled in to the 2SC state in the presence of the magnetic field.
The magnetic field would then be resolved in to a $\Qt$ part,
which would pass freely through the 2SC quark matter, and
an $\X$ part, which would become trapped in flux tubes
(Sec.~\ref{sec:fluxtube}).

The paper is structured as follows.
In Sec.~\ref{sec:type2} we calculate the Ginzburg-Landau parameter 
for 2SC quark matter, and conclude that
it is a type-II superconductor with respect to the broken $\X$ generator
as long as the pairing gap $\De$ is large enough. We estimate that
$\De\gtrsim \mu_q/16$ will suffice, which
for typical quark chemical potentials $\mu_q\sim 400\,\MeV$ 
requires $\Delta \gtrsim 25$ MeV. 
In Sec.~\ref{sec:fluxtube} we discuss the nucleation 
scenario by which the flux tubes can occur in the 
2SC superconductor, even when the magnetic field intensities 
are below the lower critical field.
We estimate the density of such flux tubes
in the hypothetical 2SC quark matter core of a neutron star.
In Sec.~\ref{sec:scattering} we calculate the Aharonov-Bohm 
scattering cross section for electrons or unpaired quarks 
interacting with color magnetic flux tubes.
Sec.~\ref{sec:relax_time} is devoted to the computation of 
relaxation time of massless electrons and unpaired blue quarks 
interacting with flux tubes via Aharonov-Bohm cross-section. 
In Sec.~\ref{sec:forces} we estimate the timescale
for expulsion of the flux tubes from the 2SC core, taking in to account
the forces on the color-magnetic flux tubes in the 2SC core
and at its boundary, but neglecting any forces on the magnetic
flux lines outside the core.
We summarize our results in Sec.~\ref{sec:conclusions}.

In our calculations we use ``Heaviside-Lorentz'' natural units with 
$\hbar = c = k_B = \eps_0 = 1$, where $k_B$ is the Boltzmann constant
and $\eps_0$ is the vacuum permittivity; the electric charge $e$
is related to the fine structure constant by $\al=e^2/(4\pi)$.

\section{Type-II color superconductivity in quark matter}
\label{sec:type2}
A superconductor is of type II if it obeys the condition
\beq
\ka \equiv \frac{\la}{\xi}>\frac{1}{\sqrt{2}},
\label{criterion}
\eeq
where $\ka$ is the Ginzburg-Landau (GL) parameter,
$\la$ is the penetration depth, and $\xi$ is the coherence length
for the superconductor.
In a system of relativistic fermions with chemical
potential $\mu$ and pairing gap $\De$, we expect
$\xi \propto 1/\De$,  $\la\propto (g\mu)^{-1}$, so $\ka\propto\De/(g\mu)$.
(In the case of 2SC quark matter the relevant broken gauge symmetry
is the ``$X$'' which is mostly color, so the coupling $g$ is 
approximately the strong coupling constant.) We therefore expect that 
2SC quark matter will be a type-II color superconductor if
the gap is sufficiently large. 

To make a more accurate determination 
we follow the approach of Bailin and Love \cite{Bailin:1983bm}
and Iida and Baym \cite{Iida:2002ev}.
We start with the effective free energy density (Ginzburg-Landau theory)
for a relativistic
BCS superconductor (Ref.~\cite{Bailin:1983bm},~(3.12))
\be 
\label{GL_functional}
{\cal F} = {\cal F}_n+\alpha \psi^*\psi+\frac{1}{2}\beta(\psi^*\psi)^2+
\gamma(\vecnabla\psi^*-2ie\vecA\psi^*)(\vecnabla\psi+2ie\vecA\psi)
+\frac{1}{2\mu_0}(\vecB-\mu_0 \vecH)^2 \ .
\ee
(We have followed Ref.~\cite{Bailin:1983bm} in writing the magnetic field
free energy in SI units; in natural units $\mu_0=1$.)
Here $\psi$ is the gap parameter; for negative $\al$ the free energy
has a minimum at $|\psi|^2=\psi_0^2$, with
penetration depth $\la$, and coherence length $\xi$ given by
\begin{equation}
\psi_0^2 = -\frac{\alpha}{\beta}, \qquad
\lambda^2 = \frac{1}{2 \gamma q_{\rm pair}^2 \vert\psi_0\vert^2 }, \qquad
\xi^2 = -\frac{\gamma}{\alpha} \ ,
\label{GL-kappa}
\end{equation}
where $q_{\rm pair}$ is the charge of the Cooper pair.
The GL parameter $\ka$ is then given by
\beq
\kappa^2 = \frac{\lambda^2}{\xi^2}
 = \frac{1}{2 q_{\rm pair}^2}\frac{\beta}{\ga^2}.
\label{kappasq}
\eeq
The coefficients in the Ginzburg-Landau 
functional are \cite{Bailin:1983bm}
\beq
\ba{rcl}
\alpha &=& \dsp  \nu \frac{\tau_{GL}}{2},      \\[2ex]
\beta  &=&  \dsp \nu \frac{7\zeta(3)}{16(\pi T_c)^2},  \\[2ex]
\gamma  &=& \dsp \frac{\beta}{6}\frac{p_F^2}{\mu^2} ,
\ea
\label{GLcoeffs-general}
\eeq
where $\tau_{GL}\equiv (T-T_c)/T_c$. The fermions have
Fermi momentum $p_F$, so the
density of states near the Fermi surface is
$\nu= N p_F\mu/\pi^2\simeq N \mu^2/\pi^2$. The parameter $N$
is a degeneracy factor that is
$1$ for a single-species system, and $2$ for the 2SC phase
(see Ref.~\cite{Bailin:1983bm}, Eq.~(4.63)).
The Ginzburg-Landau theory is most reliable for temperatures close to $T_c$,
however we will use it at $T\ll T_c$. The low-temperature gap
parameter $\De$ is related to the
critical temperature by $T_c = (e^\ga/\pi)\De$: note that $\De$
then differs from $\psi_0$ by a factor of about 1.7. 
Expressing the coefficients in terms of $\De$,
\beq
\kappa \approx \frac{32.74}{q_{\rm pair}\sqrt{N}} \frac{\De}{\mu},
\label{kappa} \ .
\eeq

We can check this result by noting that for a relativistic electronic 
superconductor, $N=1$ and $q_{\rm pair}=2e$
where $\alpha=e^2/(4\pi)\approx 1/137$. Substituting these values into
\eqn{kappa} we find $\kappa=54.043 \Delta/\mu = 95.325 T_c/\mu$,
in agreement with Ref.~\cite{Bailin:1983bm},~(3.24).

In 2SC quark matter, the degeneracy factor is $N=2$ and the charge of 
the Cooper pair is the $\X$ charge of the 2SC condensate. 
From Eqs.~\eqn{couplings} and \eqn{qc} of Sec.~\ref{sec:scattering} we find
\beq
q_{\rm pair}= q_c \eX =
\frac{g}{\sqrt{3}\cos\mixang} \approx \frac{g}{\sqrt{3}}, 
\label{qpair}
\eeq
where the mixing angle $\mixang$ is defined in Eq.~(\ref{couplings}).
We estimate the strong coupling constant $g$ by assuming that 
$\al_s=g^2/(4\pi)\approx 1$, so $g\approx 3.5$. Substituting these values into
\eqn{kappa} we find
\beq
\kappa_{\rm 2SC} \approx 11\frac{\Delta}{\mu_q} \ .
\label{kappa-2SC}
\eeq
We conclude, using \eqn{criterion}, that
2SC quark matter will be of type II if the pairing gap
is sufficiently large, $\De \gtrsim \mu_q/16$.
In quark matter we expect $\mu_q\sim 400~\MeV$, so this only requires the
2SC pairing gap to be greater than about 25~\MeV, which is
well within typical estimates~\cite{Brown:1999aq,Brown:1999yd}.
Our general conclusion agrees with that of 
Refs.~\cite{Iida:2002ev,Giannakis:2003am}
who also noted that a sufficiently large
2SC pairing gap yields a type-II superconductor. Our specific result
\eqn{kappa-2SC} differs from Eq.~(112) of Ref.~\cite{Iida:2002ev} 
by a factor of $\sqrt{2}$, but given the uncertainty in the
strong coupling constant $g$ this numerical discrepancy does not
affect our conclusion.

\section{Color-magnetic flux tubes in the 2SC phase}
\label{sec:fluxtube}

\subsection{The nucleation and density of flux tubes}
When the quark matter core of the star cools below a critical value,
a 2SC condensate forms. We expect that this happens 
before the nuclear mantle becomes superconducting
because the gap parameter for quark matter is expected to be
an order of magnitude larger than that for 
proton pairing \cite{Alford:2007xm,Dean:2002zx,Muther:2005cj,Sedrakian:2006xm}.
The electromagnetic field is then resolved in to
a $\Qt$ component and an $X$ component. The 2SC core is not a superconductor
with respect to $\Qt$, so the $\Qt$ component is undisturbed
\cite{Alford:1999pb} (on this we disagree
with Ref.~\cite{hep-ph/0012383},
which we believe imposes an incorrect boundary condition on the gluon field).
However, the core is
a superconductor with respect to the $X$ component, and we have argued above
that it may well be a type-II superconductor. The lower critical field
for the $X$-superconductivity is very high, $H_{c1} \sim 10^{17}$\,Gauss
\cite{Iida:2002ev}, and typical neutron star magnetic fields are expected
to be lower than this, but, as we now argue
(see also \cite{Alford:1999pb} and footnote [8] of Ref.~\cite{Iida:2002ev}), it
is still quite possible for the $X$-flux 
to form flux tubes threading the quark matter core.
The only way the $X$-flux could be expelled from the core is if the
transition from hot quark matter to 2SC happens smoothly
from the center of the star outwards.
However, it seems more likely that
the transition to 2SC matter will proceed by
nucleation of 2SC regions (``bubbles'') in the quark matter, 
which then grow and coalesce. The $X$-flux will 
be expelled from the 2SC bubbles, but will then be
trapped in the non-superconducting regions between the bubbles.
As the bubbles grow, these regions become smaller, concentrating the
flux there until the local field strength rises above $H_{c1}$, 
at which point the bubbles stop
growing. At this stage, the core consists of 2SC quark matter with
channels of non-superconducting quark matter running though it, carrying
the $X$-flux. If the 2SC phase is a type-II superconductor then these
channels are unstable and will fragment into flux tubes, each carrying
a single quantum of $X$-flux, with a short-range repulsion between the
flux tubes.
The fact that the average field strength was below the lower critical
field for a sphere of 2SC matter in a uniform magnetic field will now
manifest itself as an outwardly-directed
boundary force on the flux tubes at the point
where they meet the edge of the 2SC core. We will study this in
Sec.~\ref{sec:forces}.

Because the 2SC phase is a conductor with respect 
to $\Qt$ charge, it supports eddy currents which make it
very difficult for the $\Qt$ magnetic field in the 2SC core to 
change. The timescale for expulsion of the $\Qt$ magnetic field
is estimated to be longer than the age of the universe \cite{Alford:1999pb}.
Thus we are justified in 
treating the $\Qt$ magnetic field as a fixed background.

If we assume that all the $X$-flux is trapped in the manner described above, 
then the density
of flux tubes is just $B_X$, the density of magnetic $X$-flux, divided by 
$\Phi_X$, the $X$-flux of a single flux tube.
$B_X$ is obtained by projecting out the $X$-component
of the original electromagnetic flux $B$ (see \eqn{mixing}), so
$B_X=B\sin\mixang$. 
The flux quantum is 
\beq
\Phi_X = \frac{2\pi}{q_{\rm pair}} 
\approx \sqrt{\frac{3\pi}{\alpha_s}} \ ,
\label{X-quantum}
\eeq
where $q_{\rm pair}$  is the 
$X$-charge of the 2SC condensate (see \eqn{qpair} and \eqn{couplings}).
We can relate it to the flux quantum  $\Phi_0=\pi/e\approx 10.37$
for an ordinary superconductor
where the charge of the condensate is $2e$,
\beq
\Phi_X = \frac{2e}{q_{\rm pair}} \Phi_0 = 6\sin(\mixang) \Phi_0 \ .
\label{XvsPhi0}
\eeq
We conclude that
\be\label{eq:flux_number} 
n_{v} = \frac{B_X}{\Phi_X} =
\frac{1}{6} \frac{B}{\Phi_0} \ .
\ee
This is the upper limit on the flux tube density in 2SC matter.
Interestingly, as anticipated in Ref.~\cite{Blaschke:2000gm},
it only differs by a factor of
$1/6$ from the density of electromagnetic flux tubes that would result if the
core were an electromagnetic superconductor due to electron or proton pairing.
Projection on to the $X$ component reduces the magnetic flux by
a factor $\sin\mixang$, but because the $X$ fields are strongly coupled 
their flux quantum is smaller by a similar factor, so the flux tube
density ends up being independent of the mixing angle.
The actual density will depend
on details of how the transition to 2SC matter was completed.
For an internal field $B=10^{14}$\,Gauss ($2\,\MeV^2$), the maximum
flux tube density is $n_v=8.1\times 10^{19}\,{\rm cm}^{-2}$.


\subsection{Properties of the flux tube}

The thickness of the flux tubes is given by the
penetration depth for magnetic $X$-flux in the 2SC phase. This follows
from equations \eqn{kappasq} to \eqn{qpair}. Assuming $p_F\simeq \mu_q$
for relativistic quarks,
\be
\lambda = \frac{3\pi}{g\mu_q\vert\tau_{GL}\vert^{1/2}}
= (1.3\,\fm) \left(\frac{400{\rm~MeV}}{\mu_q}\right) 
\left(1-\frac{T}{T_c}\right)^{-1/2} \ .
\label{lambda-X}
\ee
The energy per unit length (tension) of the flux tube is given by
$\half {\cal E} \ln\kaX$ where ${\cal E}$ is the energy per unit length
of the magnetic flux if it were uniformly spread over an circle
of radius $\la$ (Ref.~\cite{Tinkham}, Sec.~(5.1.2)), and $\ln\ka_X$
is a factor of order 1. In Heaviside-Lorentz
natural units ${\cal E} = B^2/2$, where $B=\Phi_X/(\pi\la^2)$, so
\be 
\ep_X = \frac{\Phi_X^2}{4\pi\lambda^2}\ln \kaX 
\ee
(compare Ref.~\cite{Iida:2002ev}, Eq.~(107); 
see also Ref.~\cite{Blaschke:2000gm}).
To estimate the tension we work to lowest order in $\al$ and use
\eqn{lambda-X}, \eqn{X-quantum}, and \eqn{sinmix}.
In the low temperature limit we find
\beq
\ep_X = \frac{\mu_q^2}{3\pi}\,\ln\kaX \ .
\label{X-tension}
\eeq
Assuming that in 2SC quark matter $\mu_q$ is in the 350 to 500 MeV range,
and that the logarithmic factor is of order 1,
we conclude that the tension will be of order $60$ to $130$\,MeV/fm.


\section{Aharonov-Bohm scattering by flux tubes}
\label{sec:scattering}

The Aharonov-Bohm effect provides a remarkably strong
interaction between a charged particle and 
a flux tube containing magnetic flux.
For the simple case of a single $U(1)$
gauge group (electromagnetism),
the differential cross-section per unit length is
(see, for example, Ref.~\cite{Alford:1988sj})
\begin{equation}
 \frac{d\si}{d\vartheta} = 
\frac{\sin^2(\pi\tilde\beta)}{ 2\pi k\sin^2(\vartheta/2)},
\label{AB-scattering}
\end{equation}
where
\beq
\tilde\beta= \frac{q_p}{q_c} \ ,
\label{AB-parameter}
\eeq
where $q_p$ is the charge of the scattering particle. 
For a flux tube that arises as a topological soliton in an Abelian
Higgs model, $q_c$
is the charge of the condensate field whose winding by a phase of
$2\pi$ characterizes the flux tube; $k$ is the momentum in the plane
perpendicular to the string, and $\vartheta$ is the scattering angle.
Aharonov-Bohm scattering has several important features:
\begin{tightlist}{$\bullet$}
\item The cross-section vanishes if $\tilde\beta$ is an integer, but
is otherwise non-zero.
\item The cross section is {\em independent of the thickness
of the flux tube}: the scattering is not suppressed in the limit
where the symmetry breaking energy scale goes to infinity, and
the flux tube thickness goes to zero.
\item The cross section diverges both at 
low energy and for forward scattering.
\end{tightlist}
It is therefore of great interest to determine the values of
$\tilde\beta$ for scattering of the
fermions that are ungapped in the 2SC phase
off a flux tube containing magnetic flux
associated with the broken gauge symmetry.

\subsection{The gauge groups and charges}

\subsubsection{The light fermions}
In the 2SC phase we will focus on the $U(1)\times U(1)$
gauge group consisting of electromagnetism and the part of the
color gauge symmetry that mixes with electromagnetism. 
The relevant particles are the quarks and the electron:
\beq
\psi = (ru,gd,rd,gu,bu,bd,e^-),
\label{basis}
\eeq
where ``$ru$'' means the red up quark, etc. ``$e^-$'' is the electron.
Muons would have the same interaction as the electron, so we
do not include them separately.
In this basis, the generators of the two $U(1)$ gauge groups are
just the diagonal matrices of their electric and color charges,
\beq
\ba{rcl}
Q^\psi &=& {\rm diag}(+\twothirds,-\third,-\third,+\twothirds,
  +\twothirds,-\third,-1), \\[1ex]
T^\psi &=& \frac{1}{2\sqrt{3}}{\rm diag}(1,1,1,1,-2,-2,0).
\ea
\label{generators}
\eeq
The normalization of $Q^\psi$ is fixed by the conventional
electric charges of the particles. For $T$ we have used
the conventional normalization for generators of the $SU(3)$
color gauge group~\cite{Donoghue_Golowich_Holstein}.
The kinetic term in the lagrangian of the fermions 
is $\bar\psi\ga^\mu D_\mu \psi$, 
where the covariant derivative of the fermion fields is
\beq
D_\mu \psi = \p_\mu\psi - i e A^Q_\mu Q^\psi\psi -i g A^T_\mu T^\psi\psi \ .
\label{Dpsi-orig}
\eeq
The electromagnetic gauge coupling is $e$, and the QCD gauge coupling is $g$.
The photon gauge field is $A^Q$, and the gluon gauge field is $A^T$.
With the normalization of \eqn{generators}, 
$\alpha=e^2/4\pi=1/137$, and $\al_s=g^2/4\pi\sim 1$.

\subsubsection{The 2SC condensate}
The 2SC condensate is a diquark condensate,
\beq
\phi_{ij} = \<\psi_i C\ga_5 \psi_j\> \ ,
\eeq
where the indices $i$ and $j$ live in the color-flavor space of
\eqn{basis}.
The condensate only involves the red and green up and down quarks, so
its color-flavor structure is
\beq
\phi \propto \left(
\ba{rrrrrrr}
0 & 1 & 0 & 0 & 0 & 0 & 0 \\
1 & 0 & 0 & 0 & 0 & 0 & 0 \\
0 & 0 & 0 & -1 & 0 & 0 & 0 \\
0 & 0 & -1 & 0 & 0 & 0 & 0 \\
0 & 0 & 0 & 0 & 0 & 0 & 0 \\
0 & 0 & 0 & 0 & 0 & 0 & 0 \\
0 & 0 & 0 & 0 & 0 & 0 & 0
\ea
\right)\ .
\label{phi2SC}
\eeq
From \eqn{generators} we can see how $\phi_{ij}$, considered
as a $7\times7$ matrix in the color-flavor space
of \eqn{basis}, transforms
under an infinitesimal electromagnetism or color rotation. Each of the
quarks in the diquark feels its own color-flavor phase, so each
index $i$ and $j$ is separately transformed:
\beq
\ba{rcl}
Q^\phi \phi &=& Q^\psi\cdot \phi + \phi\cdot Q^\psi, \\[1ex]
 T^\phi \phi &=& T^\psi \cdot\phi + \phi\cdot T^\psi,
\ea
\eeq
where the ``$\cdot$'' on the right hand side signifies ordinary matrix
multiplication of the two $7\times 7$ matrices, and
we have used the fact that $Q^\psi$ and $T^\psi$ are both diagonal,
and hence symmetric.

The lagrangian of the 2SC condensate (\ie~the G-L theory) contains
the kinetic term $(D_\mu\phi)^*D^\mu\phi$, where the covariant derivative
is
\beq
D_\mu\phi = \p_\mu\phi -ieA^Q_\mu Q^\phi\phi -igA^T_\mu T^\phi\phi \ .
\eeq 
This determines the coupling of the 2SC condensate to the gauge fields.

\subsection{The broken/unbroken basis}

When the 2SC condensate $\phi$ forms, one linear combination of
$Q$ and $T$ is spontaneously broken: we will call it ``$\X$''.
The other remains unbroken: we will call it ``$\Qt$'',
\beq
\ba{rcl}
\Qt &=& Q + \eta_1 T \ , \\[1ex]
\X &=& -\eta_2 Q + T \ .
\ea
\label{gens-new}
\eeq
We determine $\eta_1$ by requiring that the 2SC condensate
be invariant under $\Qt$ gauge transformations,
\beq
\Qt^\phi\phi = 0 \ ,
\eeq
which implies that
\beq
\eta_1 = -\frac{1}{\sqrt{3}} \ .
\label{eta1}
\eeq
It is natural to work in the $(\Qt,X)$ basis rather than the $(Q,T)$
basis, so we define new  ``rotated'' gauge fields
\beq
\ba{rcl}
A^\Qt &=& \cos\mixang A^Q - \sin\mixang A^T, \\
A^\X &=&  \sin\mixang A^Q + \cos\mixang A^T,
\ea
\label{mixing}
\eeq
where the mixing angle $\mixang$
is analogous to the Weinberg angle in the
standard model which parametrizes the mixing of the hypercharge and
$W^3$ gauge bosons to yield the photon (analogous to $A^\Qt$ here) and
the $Z$ (analogous to $A^\X$ here). It is important that the mixing of the
gauge fields is expressed in terms of an angle, so it
maintains their normalization, so the gauge field kinetic
terms for $A^\Qt$ and $A^\X$ are still conventionally normalized.
In the case of the generators, which we defined in \eqn{gens-new}, the 
overall normalization is not important, since it is absorbed
in the new gauge couplings.

In the new basis, the covariant derivative of the fermions is
\beq
D_\mu\psi = \p_\mu\psi -i\eQt A^\Qt_\mu \Qt^\psi\psi 
 - i\eX A^\X_\mu \X^\psi\psi \ .
\label{Dpsi-new}
\eeq
We will determine the new gauge couplings $\eQt$ and $\eX$, and
the mixing parameters $\eta_2$ and $\mixang$, by requiring that
\eqn{Dpsi-new} be equivalent to \eqn{Dpsi-orig} for all gauge field 
configurations.

\subsection{$\X$-charges of the particles and condensate}

Flux tubes will contain magnetic $X$-flux, so to
determine the Aharonov-Bohm scattering parameter $\tilde\beta$ 
for each particle, 
we need to find the $\X$-charge of each particle, corresponding to $q_p$
in \eqn{AB-parameter}. This follows straightforwardly from 
\eqn{Dpsi-new}. We will also need to know the $\X$-charge of the 
2SC condensate, corresponding to $q_c$ in \eqn{AB-parameter}.

Requiring that \eqn{Dpsi-new} be equivalent to \eqn{Dpsi-orig} for all 
gauge field configurations, and using \eqn{eta1}, we find
\beq
\ba{rcl}
\cos\mixang &=& \dsp \frac{\sqrt{3}g}{\sqrt{e^2 + 3g^2}} \\[3ex]
\eta_2 &=&\dsp -\frac{e^2}{\sqrt{3}g^2} 
  = -\sqrt{3}\tan^2\mixang  \\[3ex]
\eQt &=& \dsp\frac{\sqrt{3}eg}{\sqrt{e^2 + 3g^2}} = e\cos\mixang \\[3ex]
\eX &=&\dsp \frac{\sqrt{3}g^2}{\sqrt{e^2 + 3g^2}} = g\cos\mixang\ .
\ea
\label{couplings}
\eeq
There is a new ``rotated'' electromagnetism, with coupling
$\eQt$, which is slightly smaller than the usual electromagnetic
gauge coupling. The charges of the fermions under this gauge group
are
\beq
\Qt = {\rm diag}( \half,-\half,-\half,\half,1,0,-1) \ .
\label{eq:Qcharges}
\eeq
This agrees with the well-known results for the 2SC phase
\cite{Alford:2007xm}.

The action of the $\X$-charge matrix on the 2SC condensate determines
the $\X$-charge $q_c$ of the condensate, in units of $\eX$; 
from \eqn{phi2SC}, \eqn{gens-new},
and \eqn{couplings},
\beq
\ba{rcl}
X \phi + \phi X  &=& q_c \phi , \\[3ex]
\hbox{where}\quad
   q_c &=&\dsp  \frac{1}{\sqrt{3}}\Bigl( 1 + \frac{e^2}{3g^2} \Bigr) 
 = \frac{1}{\sqrt{3}\cos^2\mixang}.
\ea
\label{qc}
\eeq
The $X$-charge matrix of the fermions is
\beq
\ba{rcl@{}l@{\;}l}
 X &=& 
\dsp\frac{1}{\sqrt{3}} {\rm diag}(&\dsp
  \half+2\tan^2\mixang,
   &\dsp \half-\tan^2\mixang, \\[1ex]
&& &\dsp \half-\tan^2\mixang, 
   &\dsp\half+2\tan^2\mixang, \\[1ex]
&& &\dsp  -1+2\tan^2\mixang, 
   &\dsp-1-\tan^2\mixang, \\[1ex]
&& &\dsp -3\tan^2\mixang ).
\ea
\eeq
Dividing by $q_c$ \eqn{qc} we find the Aharonov-Bohm $\tilde\beta$-factors
of the fermions, in the basis defined by \eqn{basis},
\beq 
\ba{r@{}rrr}
\tilde\beta^\psi = {\rm diag}\Bigl(
  &\dsp \frac{1}{2}+\frac{3}{2}\sin^2\mixang,  
  &\dsp \frac{1}{2}-\frac{3}{2}\sin^2\mixang, \\[2ex]
  &\dsp \frac{1}{2}-\frac{3}{2}\sin^2\mixang,
  &\dsp \frac{1}{2}+\frac{3}{2}\sin^2\mixang, \\[2ex]
  &\dsp -1+3\sin^2\mixang, 
  &\dsp -1,  
  &\dsp -3\sin^2\mixang \Bigr) \ .
\ea
\eeq
Expanding in powers of $e^2$ (since $e\ll g$), we find
\beq
\sin^2(\mixang) \approx \frac{\al}{3\al_s}
\label{sinmix}
\eeq
so to lowest order in $\al$,
\beq 
\ba{r@{}rrr}
\tilde\beta^\psi = {\rm diag}\Bigl(
  &\dsp \frac{1}{2}+\frac{\al}{2\al_s},  
  &\dsp \frac{1}{2}-\frac{\al}{2\al_s}, \\[2ex]
  &\dsp \frac{1}{2}-\frac{\al}{2\al_s}, 
  &\dsp \frac{1}{2}+\frac{\al}{2\al_s}, \\[2ex]
  &\dsp -1+\frac{\al}{\al_s}, 
  &\dsp -1,  
  &\dsp -\frac{\al}{\al_s}  \Bigr) \ .
\ea
\label{ABfactor-approx}
\eeq
We conclude that the gapped quarks have 
$\tilde\beta$ close to $\half$, which means
that they have near-maximal Aharonov-Bohm interactions with an $X$-flux tube.
Among the lighter (and hence more phenomenologically relevant) fermions,
the $\Qt$-neutral $bd$ has zero Aharonov-Bohm interaction with the flux tube,
while the $bu$ and electron have the same Aharonov-Bohm factor
\beq
\sin(\pi\tilde\beta^{bu})=\sin(\pi\tilde\beta^{e})
\approx -\pi \frac{\alpha}{\alpha_s} \ .
\label{ABfactor-bu}
\eeq

\section{Relaxation via scattering off flux tubes}
\label{sec:relax_time}

\subsection{Relaxation time calculation}

In this section we compute the characteristic timescale for
a perturbation from equilibrium to relax away
due to scattering of the fermions off the color 
magnetic flux tubes. This relaxation time is a measure of the
mean free time between collisions of the fermions with
the flux tubes, so we will also refer to it as a collision time.
Our calculation applies equally to electrons
and the unpaired component of blue colored quarks in the 2SC phase,
the key difference being $\tilde\beta$ factors in the cross-section. 
The Boltzmann kinetic equation for blue-quark/electron distribution 
function  $f(\vecp, t)$ is 
\bea\label{eq:Boltzmann}
\frac{\partial f(\vecp, t)}{\partial t} 
&=&\frac{2\pi N_v}{V}
\int\!\!\frac{d^3p'}{(2\pi)^3} \Biggl\{
 W(\vecp;\vecp')f(\vecp',t)\left[1-f(\vecp,t)\right]
\nonumber\\
&-&W(\vecp';\vecp)f(\vecp,t)\left[1-f(\vecp',t)\right]
\Biggr\}\delta(\ep(\vecp)-\ep(\vecp') ) ,
\eea
where $N_v$ is the number of flux tubes, $V$ is the volume,
and $W(\vecp';\vecp)$ is the transition probability between
the states described by momenta $\vecp$ and $\vecp'$. 
Time-reversal symmetry implies  $W(\vecp;\vecp')=W(\vecp';\vecp)$.
In equilibrium the fermion distribution function is given
by the Fermi-Dirac distribution function 
\be\label{eq:Fermi-Dirac}
f_0(\vecp)= \frac{1}{1+\exp[(p-\mu_i)/T]}
\ee
where $T$ is the temperature and $\mu_i$ is 
the chemical potential of blue quarks ($i=b$) and 
electrons $(i=e)$. To solve the Boltzmann
equation we shall apply the variational method, where the perturbations
from equilibrium are described by variational trial 
functions whose functional form is dictated by the form of 
applied perturbation~\cite{Flowers:1976,Flowers:1979}. 
The resulting 
transport coefficients are lower bounds on their exact values.
The number of adjustable trial functions, which are used to maximize
the entropy production via scattering, could be large. In the following
we shall use one linear function $\phi$, in which case there is no need 
for variation, since the variational parameter cancels out. It should 
be kept in mind that the resulting transport coefficients are still lower 
bounds on their exact values.  For small perturbations
from equilibrium the Boltzmann equation can be linearized by writing
$f(\vecp,t) = f_0(\vecp)+\delta f(\vecp,t),$ where the (small) 
perturbation from the Fermi-Dirac form (\ref{eq:Fermi-Dirac}) is 
\be\label{eq:perturb}
\delta f(\vecp,t) = -\frac{df_0(\vecp)}{d\ep(\vecp)}\phi(\vecp,t),
\ee
where $\phi(\vecp,t)$ is the trial function. The linearized Boltzmann
equation then reads
\bea\label{eq:Boltzmann_linear}
 - \frac{\partial\phi(\vecp,t)}{\partial t}f_0(\vecp)\left[1-f_0(\vecp)\right] 
&=&\frac{2\pi N_v}{V}
\int\!\!\frac{d^3p'}{(2\pi)^3}\left[\phi(\vecp,t)-\phi(\vecp',t)\right] 
\nonumber\\
&\times&W(\vecp;\vecp')f_0(\vecp')\left[1-f_0(\vecp)\right] 
\delta(\ep(\vecp)-\ep(\vecp')).
\eea
To obtain this form of the kinetic equation 
we used the detailed balance conditions
$
 f_0(\vecp')\left[1-f_0(\vecp)\right] 
-  f_0(\vecp)\left[1-f_0(\vecp')\right] = 0,
$  and 
$
{df_0(\vecp)}/{d\ep(\vecp)} = {df_0(\vecp')}/{d\ep(\vecp')} .
$
It is convenient to work with the Laplace transformed trial function 
\be 
\phi(\vecp,t) = \int ds e^{-st}\phi(\vecp,s).
\ee
Upon Laplace transforming  Eq.~(\ref{eq:Boltzmann_linear}) we find
\bea\label{eq:motion}
s \phi(\vecp,s)f_0(\vecp)\left[1-f_0(\vecp)\right]  
&=&\frac{2\pi N_v}{V} \int\!\!\frac{d^3p'}{(2\pi)^3}
\left[\phi(\vecp,s)-\phi(\vecp',s)\right]\nonumber\\
&\times&W(\vecp;\vecp')
f_0(\vecp')\left[1-f_0(\vecp)\right] \delta(\ep(\vecp)-\ep(\vecp')).
\eea
To define a characteristic relaxation rate we assume that the 
trial function can be written as 
\be
\phi(\vecp,s) = \phi(\vecp)\delta(s-s_0),
\ee
in which case Eq.~(\ref{eq:motion}) becomes 
\bea\label{eq:motion2}
s_0 \phi(\vecp)f_0(\vecp)\left[1-f_0(\vecp)\right]  
&=&\frac{2\pi N_v}{V} \int\!\!\frac{d^3p'}{(2\pi)^3}
\left[\phi(\vecp)-\phi(\vecp')\right]\nonumber\\
&\times&W(\vecp;\vecp')
f_0(\vecp')\left[1-f_0(\vecp)\right] \delta(\ep(\vecp)-\ep(\vecp')),
\eea
where the perturbation functions are now independent of $s$. 
We can identify $s_0$ with the relaxation rate (\ie the
inverse of the relaxation time)  by comparing the computed kinetic 
coefficients with standard expressions for transport coefficients, 
\eg, the electrical conductivity  with the Drude formula.

To formulate the variational principle~\cite{Ziman} 
we write Eq.~(\ref{eq:motion2}) in the compact form
\be\label{eq:X} 
X(\vecp) 
= \int  \left[\phi(\vecp)-\phi(\vecp')\right] P(\vecp,\vecp')d^3p',
\ee
where $X(\vecp)$ stands for the left-hand side of Eq.~(\ref{eq:motion2});
the scattering operator $P(\vecp,\vecp')$ is easily read-off 
from the kernel on the right-hand side 
of Eq.~(\ref{eq:motion2}). Since the  factor $f_0(\vecp)
\left[1-f_0(\vecp)\right]$ and the transition probability are positive 
definite, the operator $P(\vecp,\vecp')$ is positive definite. 
Further it is linear and self-adjoint (symmetric).
Following ref.~\cite{Ziman} we define an inner product 
\be
\langle\phi , \psi \rangle \equiv \int \phi(\vecp)\psi(\vecp) d\vecp,
\ee
in terms of which
\be\label{eq:12}
\langle\phi , P\psi \rangle \equiv
\frac{1}{2}\int d\vecp\int d\vecp' [\phi(\vecp)
-\phi(\vecp')]P(\vecp,\vecp')[\psi(\vecp)-\psi(\vecp')].
\ee
The variational principle states that the expression 
\be \label{var_principle}
\langle\phi , X\rangle = \langle\phi , P\phi\rangle 
\ee
attains its maximum for the {\em exact} value
$\phi_{\rm ex}$, which satisfies Eq.~(\ref{eq:X}); 
for any other trial function $\phi$, that satisfies
Eq.~(\ref{var_principle}),   
$\langle\phi , P\phi\rangle\le 
\langle\phi_{\rm ex} , P\phi_{\rm ex}\rangle$. 
Explicitly, Eq.~(\ref{var_principle}) reads
\bea\label{eq:s1}
s_0 \int\frac{d^3p}{(2\pi)^3}
\phi(\vecp,s)^2f_0(\vecp)\left[1-f_0(\vecp)\right]
&=&\frac{2\pi N_v}{ V}
\int\!\!\frac{d^3p}{(2\pi)^3}
\int\!\!\frac{d^3p'}{(2\pi)^3} \frac{1}{2}
\left[\phi(\vecp,s)-\phi(\vecp',s)\right]^2\nonumber\\
&&W(\vecp;\vecp')
f_0(\vecp')\left[1-f_0(\vecp)\right] \delta(\ep(\vecp)-\ep(\vecp')).
\eea
It is also straightforward to 
check that the variation of Eq.~(\ref{eq:s1}) leads us back 
to ``equation of motion''
(\ref{eq:motion2}). From Eq.~(\ref{eq:s1}) we  
obtain the variational relaxation rate 
\bea\label{eq:s1bis}
s_0 &= &\frac{2\pi N_v}{ V{\cal D}}
\int\!\!\frac{d^3p}{(2\pi)^3}
\int\!\!\frac{d^3p'}{(2\pi)^3} \frac{1}{2}
\left[\phi(\vecp)-\phi(\vecp')\right]^2W(\vecp;\vecp')
f_0(\vecp')\left[1-f_0(\vecp)\right] \delta(\ep(\vecp)-\ep(\vecp')),
\nonumber\\
\eea
where 
\be \label{eq:calD}
{\cal D} = \int\frac{d^3p}{(2\pi)^3}
\phi(\vecp)^2f_0(\vecp)\left[1-f_0(\vecp)\right].  
\ee
The exact relaxation rate $s \ge s_0$.
We specify 
the form of the trial function appropriate to the problem at hand, 
which is the relaxation of uniform blue-quark/electron velocity 
$\vecv$ on a flux tube, as
\be \label{eq:trial2}
\phi(\vecp) = \vecp \cdot \vecv ~C(p^2),
\ee
where $C(p^2)$ is the scalar part of the trial function. 
In the following we will adopt the simple choice $C(p^2)=1$. 
The differential transition probability can be obtained from 
the Aharonov-Bohm scattering cross-section, Eq.~(\ref{AB-scattering}),
and is given by (for details see Appendix~\ref{app:corss_section})
\be\label{eq:diff_probability}
dW = 2\pi\delta(\ep'-\ep) 2\pi\delta(p_z-p_z')
\frac{4 L\sin^2(\pi\tilde\beta)}{\sin^2(\phi/2)} \frac{1}{2\ep V}
 \frac{d^3p'}{(2\pi)^32\ep'},
\ee
where the initial and final state momenta and energies, $p$ and $\ep$, are 
unprimed and primed respectively (in this section 
we use $\phi$ for the scattering
angle, as opposed to $\vartheta$ in Sec.~\ref{sec:scattering}).
Here we have used the cylindrical coordinates coaxial with the flux tube
to write $d^3p = p_{\perp}dp_{\perp} d \phi d p_z$. Combining 
Eqs.~(\ref{eq:s1bis}) and (\ref{eq:diff_probability}) and carrying
out the phase space integrals (the details are given in  
Appendix~\ref{app:phase_space}) we obtain, to lowest order in the 
low-temperature expansion, 
\bea\label{eq:s5}
s_0&=& \frac{p_{Fi}^3 v^2n_v T}{6\pi^2 {\cal D}}
~\sin^2(\pi\tilde\beta),
\eea
where $p_{Fi}$ is the blue-quark/electron Fermi momentum, 
$n_v$ is the density of flux tubes. Eq.~(\ref{eq:calD}) 
with the trial function (\ref{eq:trial2})
can be computed in the low-temperature limit by approximating 
${df(\vecp)}/{d\ep(\vecp)} \simeq 
\delta(\ep(\vecp)-\mu_i)$ to obtain
${\cal D} = {p_{Fi}^4  v^2T}/{6 \pi^2}$,
where $v$ is the fermion fluid velocity \eqn{eq:trial2}.
The relaxation rate for particles of species $i$
scattering off flux tubes of area density $n_v$ is then given by
\be
\tau^{-1}_{if} \equiv s_0= \frac{n_v}{p_{Fi}}  \sin^2(\pi\tilde\beta_i) \ .
\label{tauinv-flux}
\ee
It easy to understand the final result (\ref{tauinv-flux}). It is of the
standard form for classical gases $\tau^{-1}=c n \si$, 
where $c=1$ is the speed of
the particles, $n=n_v$ is the density of scattering centers,
and $\si\propto \sin^2(\pi\tilde\beta)/p_F$ is the cross section for
Aharonov-Bohm scattering. Eq.~\eqn{tauinv-flux} is relevant for thermal
relaxation of the gapless fermion species in the 2SC phase.
One of these, the blue down quark, has no A-B interaction with the flux tubes
($\tilde\beta=0$). The other two, the electron and blue up quark,
have identical A-B factors \eqn{ABfactor-bu}
although their Fermi momenta are different.

\subsection{Comparison with Coulomb scattering}
\label{sec:Coulomb}

To find out whether scattering off flux tubes is likely to be an important
source of relaxation, and hence a significant contributor to transport 
properties, it is useful to compare Eq.~\eqn{tauinv-flux} with the
collision time for screened Coulomb scattering via exchange of $\Qt$ photons.

The 2SC phase is a $\Qt$-conductor, with two species of gapless
charged fermions: the $bu$ quarks (with $\Qt$-charge +1 and chemical
potential $\approx\mu$) and the electrons (with $\Qt$-charge +1 and
chemical potential $\approx\mu_e$). There may also be muons,
but their Fermi momentum will be much smaller. 
As mentioned in the introduction, the red and green quarks 
will be confined to bound states
whose mass is expected to be of order 10\,MeV \cite{Rischke:2000cn}, 
so they play no role in transport at neutron star temperatures.
Since $\mu>\mu_e$, the $bu$ quarks
are more numerous than the electrons
and have a larger phase space near their Fermi
surface, so they will make the largest contribution to the collision time.

The Coulomb collision time depends on the 
in-medium photon spectrum, which will be affected by 
Debye screening and Landau damping arising from
the presence of
gapless charged excitations, dominantly the  $bu$ quarks because of their
larger phase space. A simple estimate can 
be obtained by assuming that the dispersion relation is dominated 
by a plasmon pole.  The plasma frequency $\om_p$ is given by
\beq
\omega_p^2  = \frac{\alQt n_q}{\mu_q} = \frac{4}{3\pi^2}\alQt\mu_q^2
\eeq
where $\alQt=\eQt^2/(4\pi)$ is the fine structure constant for the
``rotated'' $\Qt$ electromagnetism \eqn{couplings}.
The collision frequency
is given by (see Eq.~(10),(12),(18) of \cite{Shternin:2006uq}),
\be
\tau^{-1}_{qq} = \frac{8 \zeta(3)\mu_{q}^2}{\pi^3\omega_p^2} \alQt^2\, T
= \frac{6\zeta(3)}{\pi^2}\alQt \, T
\label{eq:qq_relax}
\ee
where $\zeta(3) = 1.202$.  This result is valid for $T\ll\omega_p$, 
which is the relevant regime for neutron stars since $\mu_q$ is in the
$400\,\MeV$ range. Eq.~\eqn{eq:qq_relax} is analogous to
Ref.~\cite{Heiselberg:1993cr}'s Eq.~(62) for the thermal conduction
timescale, with electromagnetic interactions (so their $\al_s$ is replaced
by $\alQt$) and a different number of quark species.
The quark-quark Coulomb collision frequency \eqn{eq:qq_relax}
is proportional to temperature $T$ whereas the particle-flux-tube
collision frequency is independent of temperature. We can therefore
define a temperature $T_f$ below which flux tubes dominate the
relaxation of deviations from thermal equilibrium.
From \eqn{eq:qq_relax} and \eqn{tauinv-flux} we find
\beq
T_f = 
\frac{\pi^2}{6\zeta(3)} 
\frac{\sin^2(\pi\beta_{bu})}{\alQt}
\frac{n_v}{\mu_q} \ .
\eeq
To make a numerical estimate 
we assume that the 2SC core contains the maximum flux tube density
given by Eq.~(\ref{eq:flux_number}), and that $\al_s\approx 1$.
Using \eqn{ABfactor-bu} and the fact that $\alQt\approx\al$, we find
\beq
T_f \approx (9\times 10^4\,K)  
  \Bigl(\frac{B}{10^{14}\,{\rm G}}\Bigr)
  \Bigl( \frac{400\,\MeV}{\mu_q} \Bigr) \ .
\label{T-flux_domination}
\eeq
We conclude that for reasonable values
of the magnetic field, only at very low temperatures is
Aharonov-Bohm scattering
off flux tubes likely to be an important source of 
{\em thermal} relaxation.
However, it is important to note that the thermal relaxation timescale
is not the only one that is relevant to transport. There is also
the viscous relaxation rate (Ref.~\cite{Heiselberg:1993cr}, Eq.~(51))
and the momentum relaxation rate 
(Ref.~\cite{Heiselberg:1993cr}, Eq.~(32)) both of which have a much stronger
($\propto T^{5/3}$) suppression at low temperatures. 
We defer a full discussion
of transport in the 2SC phase to later work.

\section{Forces on the flux tubes}
\label{sec:forces}

We argued in Sec.~\ref{sec:fluxtube} that even if the magnetic field
in the core of the star is below the lower critical field, color magnetic
flux tubes will still be produced in the transition to the 2SC phase.
In this section we study the forces on those flux tubes, and 
estimate the timescale for their expulsion from the 2SC core.
For this initial estimate we take in to account only the forces
on the flux tubes within the 2SC core, or at its boundary.
Depending on the nature of the material surrounding the core
there may be additional forces, and these may modify the expulsion
time in a way that would have to be calculated on a case-by-case basis.



The velocity of the flux tube is $\vecv_L$, the velocity of the
normal fluid is $\vecv_N$, and the velocity of the 2SC condensate is
$\vecv_S$.
The forces we consider are mutual friction (``mf''), 
the non-dissipative (lifting) 
Magnus-Lorentz force (``ML''), the Iordanskii force (``Iord''),
forces arising from zero modes (``zm''),
and boundary forces (``bf'') at the quark-hadronic boundary.
We assume that local magneto-hydrostatic-gravitational
equilibrium is established quickly after the transition to the 2SC phase, 
so there is no additional buoyancy force 
\cite{1985SvAL...11...80M,2009MNRAS.397.1027J}.
We note that there may be additional forces due to density-dependence of
the 2SC pairing gap \cite{Hsu:1999rf}, but we do not include these
since there is as yet no reliable estimate of the density dependence.
The equation of motion of a flux tube then has the form 
\be\label{eq:dynamics} 
m_V \frac{d\vecv_L}{dt} = 
   \vecf_{\rm mf}  
 + \vecf_{\rm ML} 
 + \vecf_{\rm Iord}
 + \vecf_{\rm zm}
 + \vecf_{\rm bf} \ ,
\ee
where $m_V$ is the effective mass of a flux tube 
per unit length and $\vecf$ is a force per unit length. 
The boundary forces tend to pull the flux tube in a radial direction,
expelling it from the 2SC core. This is resisted by
the combination of the other forces.

In our calculations we will assume
that the flux tubes are straight. A bent flux tube will feel
an additional restoring force determined by its tension.

\subsection{The background $\Qt$ magnetic field}
\label{sec:Qt-field}

In our calculations we will neglect the effect of the $\Qt$ magnetic
field $B_{\Qt}$ that penetrates the 2SC core. Because of this field,
$\Qt$-charged particles, including the $bu$ quarks and electrons,
will feel a Lorentz force. This will have a significant
effect on the behavior of the normal fluid of quarks and electrons
when the cyclotron frequency $\om_c$ becomes larger than the inverse of the
characteristic time for equilibration, 
which as we argued in Sec.~\ref{sec:Coulomb}, is
the quark-quark collision time $\tau_{qq}$ \eqn{eq:qq_relax}. 
The dominant component
of the fluid is the $bu$ quarks, with $\Qt$-charge $e^\Qt\approx e$,
and $B_{\Qt}\approx B$, so
\beq
\om_c = \frac{eB}{p_F} \ ,
\eeq
and we can neglect the effects of the
magnetic field on transport when $\om_c\tau_{qq} \ll 1$, where
\beq
\om_c\tau_{qq} =
\frac{2 \pi^3}{3 \zeta(3)} \frac{1}{\sqrt{4\pi\al}}
\frac{B}{\mu_q T} 
= 0.32 \Bigl(\frac{B}{10^{12}\,\G}\Bigr)
\Bigl(\frac{10^8\,\K}{T}\Bigr)
\Bigl(\frac{400\,\MeV}{\mu_q}\Bigr) \ .
\label{omegac*tau}
\eeq
We conclude that only for high magnetic fields (above $10^{12}$\,G)
or low temperatures (below $10^8$\,K) might the magnetic field affect
thermal relaxation. We defer a discussion of this regime to future work.

\subsection{Mutual friction}

Mutual friction is a frictional force on a flux tube arising from
its Aharonov-Bohm interaction with the normal fluid of
gapless particles through which it is moving.
Consider a vortex moving relative to the normal fluid
with velocity $\vecu=\vecv_L-\vecv_N$.
In the relaxation time approximation
\be 
\vecf_{\rm mf} = \frac{\tau^{-1}_{if}}{n_v} 
  \int\frac{d^3p}{(2\pi)^3} \, \vecp\, f_0(p,\vecu),
\ee
where $\tau^{-1}_{if}$ is the collision rate between fermions of
species $i$ and flux tubes \eqn{tauinv-flux}.

We will assume that the blue up quarks dominate the friction.
This is reasonable because the blue down quarks
have no $X$ charge and hence no Aharonov-Bohm interaction with
the flux tube, and the electron Fermi momentum
is smaller than that of the blue quarks.
The equilibrium Fermi-Dirac thermal distribution of the
quarks is
\be 
f_0(p,\vecu) = \{ \exp[(\ep-\mu_i + \vecp\cdot \vecu)/T]+1 \}^{-1} \ ,
\ee
where the $\vecp\cdot \vecu$ term is a correction due to the motion of the 
vortex relative to the thermal bath with velocity $\vecu$.
We will compute the force to linear order in $\vecu$.
The leading contribution arises at the first order in velocity
\be 
\vecf_{\rm mf} = \frac{\tau^{-1}_{if}}{n_v} 
  \int\frac{d^3p}{(2\pi)^3} \vecp (\vecp\cdot \vecu) 
\frac{\partial f_0(\ep)}{\partial \ep} = \eta\vecu.
\label{eta-defn}
\ee
Carrying out the integral and using \eqn{tauinv-flux} we obtain 
the mutual friction drag coefficient
\be 
\eta = \frac{p_{Fi}n_i \tau^{-1}_{if}}{n_v}
= n_i \sin^2(\pi\tilde\beta_i)
\label{eq:eta}
\ee
where $n_i$ is the fermion density and $\tilde\beta_i$ is their 
Aharonov-Bohm factor \eqn{ABfactor-approx}.
As one would expect, the friction coefficient is independent 
of the magnetic field (i.e.~the density of flux tubes). It is
proportional to the fermion density, so, as noted above, the $bu$ quark
contribution will dominate the electron contribution.

\subsection{Boundary forces}

\begin{figure}
\includegraphics[scale=0.6]{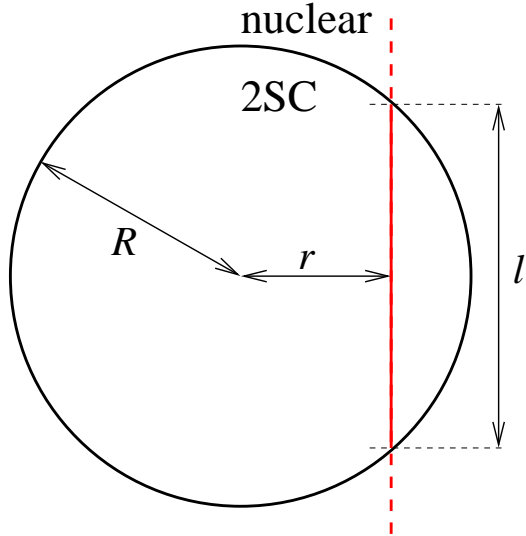}
\caption{
A straight flux tube of length $l$ passing through a 2SC neutron star core
of radius $R$, at distance $r$ from the center of the star.
There is a boundary force where it reaches the 
edge of the 2SC core.
}
\label{fig:boundary}
\end{figure}

Next we wish to calculate the force exerted on the
flux tube at the point where it reaches the interface between
the 2SC quark matter core and the nuclear mantle of a neutron star
(Fig.~\ref{fig:boundary}). When the
$X$-magnetic flux tube reaches the edge of the 2SC core
it combines with the $\Qt$ magnetic flux in the core
to re-constitute the ordinary magnetic field from which
it was originally formed. The form in which the flux
continues through the nuclear mantle, and hence the boundary
energy, may therefore be influenced by the state of the nuclear matter. 
In this analysis we will include only the forces arising from the
contribution due to the 2SC core itself. We briefly discuss other contributions,
but a proper analysis including them would have to be done in the
context of a specific model of the whole neutron star and the properties
of all regions within it.

The outward force per unit length on the flux tube is 
(see Fig.~\ref{fig:boundary})
\beq
f_b = \frac{1}{l}\frac{dE}{dr} 
 = \frac{r}{R^2-r^2}\ep_X \ .
\label{fb-full}
\eeq
Here $E$ should be the total energy of magnetic flux inside 
and outside the core, but we neglect the outside contribution;
$\ep_X$ is the energy per length of $X$ flux tubes.
Then from \eqn{X-tension},
\beq
f_b \approx \frac{r}{R^2-r^2} \frac{\mu_q^2}{3\pi}\ln\kaX \ .
\label{fbmax}
\eeq

Taking in to account the energy of the magnetic  field outside the core
will reduce the right hand side of \eqn{fb-full}, and weaken the
outward force on the flux tube.
We now discuss the magnitude of
such terms in various cases.

If the nuclear mantle is a type-II superconductor, the
magnetic field penetrates
the nuclear mantle in the form of Abrikosov
flux tubes (dashed line in Fig.~\ref{fig:boundary}).
From \eqn{eq:flux_number} we know that each $X$ flux tube
will spawn 6 Abrikosov flux tubes in the nuclear matter,
each of which has energy per unit length
\be 
\ep_{\rm nuc} = \frac{\Phi_0^2}{4\pi\lambda_{\rm nuc}^2}
  \ln \kappa_{\rm nuc} \ .
\ee
$\Phi_0=\pi/e\approx 10.37$, so if we assume that the logarithmic factor
is of order 1 then
for $\lambda_{\rm nuc}$ in the 50 to 100~fm range, $\ep_{\rm nuc}$ is in
the 0.2 to 0.7 MeV/fm range. This means that even when multiplied by
a factor of 6,  $\ep_{\rm nuc}$ is small in comparison
with the tension of the $X$ flux tube, which is
greater than 10\,MeV/fm \eqn{X-tension}, so \eqn{fbmax} is still a good
estimate of the boundary force.
Of course, in a type-II nuclear mantle there may be other forces,
for example if it is also a superfluid there may be entanglement of
Abrikosov flux tubes with superfluid vortices, but we neglect those here
because they depend on details of the nuclear mantle.

If there is no Cooper pairing of the protons then the nuclear matter is
a conductor. In this case the energy gained from shortening the
$X$ flux tube is counteracted by the field energy of the
magnetic field it connects to in the nuclear matter mantle.
The criterion for the tension of the flux tube to dominate is the same
as the criterion for the magnetic field to be below its lower critical
value. Since neutron star magnetic fields are well below
the lower critical field for the 2SC phase, we can assume that
the 2SC flux tube tension will dominate and
we can use \eqn{fbmax} again. The only complication is that 
conducting nuclear matter supports eddy currents which will resist
any change in the magnetic field in the nuclear mantle. This may
make it much harder to move the $X$-flux tubes in the 2SC core.
Again, we do not attempt to include such forces that depend on details
of the constitution of the nuclear mantle.

If the nuclear mantle were a type-I proton superconductor 
\cite{Sedrakian:2004yq,Alford:2005ku,Alford:2007np,
Charbonneau:2007db} then
$X$ flux tubes in the 2SC core would connect to
non-superconducting domains in the nuclear mantle
\cite{Sedrakian:2004yq,Charbonneau:2007db}.
In this case
we cannot compute the boundary force because the domain structure of
the type-I proton superconductor is not known; the possible (layered,
cylindrical, etc) structures in type-I superconductors essentially
depend on the history of the nucleation of the superconducting phase.

\subsection{Magnus-Lorentz force}
\label{sec:ML}

The Magnus-Lorentz force is a non-dissipative force,
directed orthogonally to the flux tube velocity, that 
arises from the superposition of the winding ``flow'' of the
2SC order parameter around the flux tube and the
background flow of the charged superfluid of 
fermions \cite{PhysRev.140.A1197,1991ApJ...380..530M,PhysRevB.55.485}.
(There is controversy about this in the literature; for example,
Jones \cite{1991MNRAS.253..279J,2009MNRAS.397.1027J} has suggested that
this is cancelled by another contribution from ungapped fermions. 
Pending a definitive resolution of this disagreement we will
use the standard form of the Magnus-Lorentz force.)

The Magnus-Lorentz force per unit length on a flux tube is
\begin{equation}
\label{eq:MLforce}
\vecf_{\rm ML} = -(\vecj_X \times \hat n \Phi_X ) \ ,
\end{equation}
where $\Phi_X$ is the $X$-flux through the
flux tube \eqn{X-quantum}, $\hat n$ is a unit vector
pointing along the flux tube, and
$\vecj_X$ is the current of $X$ charge seen by the flux tube,
arising from the $X$ charge density $\rho_X$
of the 2SC condensate, moving relative to the flux tube
\beq
\vecj_X = \rho_X (\vecv_S-\vecv_L) \ .
\eeq
We can write $\rho_X= q_{\rm pair} n_s/2$ where $n_s$ is the density
of quarks in the condensate. Since there are 4 quark species in the
condensate, and at low temperature all fermions are part of the
condensate,
\beq
\ba{rcl}
\vecf_{\rm ML} &=&\dsp -\rho (\vecv_S-\vecv_L) \times \hat n  \ , \\[2ex]
\rho &\equiv&\dsp \rho_X \Phi_X = \pi n_s = \frac{4\mu^3}{3\pi} \ .
\ea
\label{ML-final}
\eeq
Note that the charge of the Cooper pairs cancels in this expression.


\subsection{Iordanskii force}
\label{sec:Iordanskii}

The mutual friction force described above is the force on the
flux tube in the longitudinal direction (i.e.~parallel to
its velocity relative to the normal fluid of unpaired quarks), 
due to Aharonov-Bohm scattering of the unpaired quarks.
The Iordanskii force is the transverse component of 
that same force \cite{PhysRevB.55.485},
\begin{equation}
\vecf_{\rm Iord} =  D'\, (\vecv_L-\vecv_N) \times \hat n \ .
\label{Iordanskii}
\end{equation}
The transverse Aharonov-Bohm scattering cross-section for $bu$ quarks
off the flux tube is 
$\si_\perp = -k^{-1}\sin(2\pi \tilde\beta^{bu})$
(Ref.~\cite{PhysRevB.55.485}, Eq.~(64)) and, as in the
case of the longitudinal Aharonov-Bohm force, one expects
the force per unit length to be proportional to the
fermion density, so we expect
$D'\approx \sin(2\pi \tilde\beta^{bu}) \mu_q^3
\approx \al\mu_q^3$
(see Ref.~\cite{PhysRevB.55.485}, after Eq.~(69)).
This rough estimate is sufficient to argue
that the Iordanskii force can be neglected. 
Basically, the Aharonov-Bohm forces are suppressed by powers of
$\al$ arising from the Aharonov-Bohm factor of the $bu$ quarks
\eqn{ABfactor-bu}. 
In the case of the Iordanskii (transverse) component, we will see
that this makes it subleading relative to the Magnus-Lorentz force,
which also acts perpendicular to the flux tube's velocity.
In the case of the longitudinal component, there is no larger
force parallel to the velocity, so the Aharonov-Bohm force is
the dominant contribution to mutual friction.

\subsection{Zero-mode force}
\label{sec:zero-mode}

The frictional force on a flux tube
due to scattering of zero modes localized inside the flux tube
off gapless fermions in the bulk 
\cite{PhysRevLett.77.4687,Kopnin:2002}
has been calculated for
proton flux tubes in nuclear matter \cite{2009MNRAS.397.1027J}.
At low temperatures, we expect the frictional force on a 2SC flux
tube to be
\beq
\vecf_{\parallel} = - \frac{C}{\omega_0\tau_c}  (\vecv_L-\vecv_N) \ ,
\label{zm-force}
\eeq
where, generalizing from nonrelativistic protons to relativistic
quarks,
\bea
C &=& \pi n_q \tanh(\Delta/2T) \sim \mu_q^3 \ ,  \label{Cvalue} \\
\om_0 &\sim& \Delta^2/\mu_q \label{omega0} \ ,  \\ 
\tau_c &\sim& \mu_q^{2/3}T^{-5/3} \label{tau-c} \ .
\eea
Eq.~\eqn{Cvalue} follows from Ref.~\cite{2009MNRAS.397.1027J} Eq.~(7),
and the fact that 2SC pairing gap $\De$ is expected to be
much bigger than typical neutron star temperatures.
Eq.~\eqn{omega0} follows from Ref.~\cite{2009MNRAS.397.1027J} Eq.~(1),
assuming, following Ref.~\cite{2009MNRAS.397.1027J}, 
that the typical transverse momentum of the population
of zero modes is of the same order as the Fermi momentum of the quarks.
Eq.~\eqn{tau-c} is obtained by, as in Ref.~\cite{2009MNRAS.397.1027J},
assuming that scattering involving the zero modes has the same
relaxation time as quark-quark scattering in a non-superconducting
medium (i.e.~as if the flux tube core were infinitely large).
We can then use the continuum quark-quark momentum relaxation time $\tau_s$
from gluon exchange in a cold quark-gluon plasma 
(Ref.~\cite{Heiselberg:1993cr},~eqn~(28)) 
as a crude estimate of
the relaxation time $\tau_c$ for momentum transfer between
bulk gapless quarks and zero modes inside the flux tube.

Comparing \eqn{zm-force} with \eqn{eta-defn} and \eqn{ABfactor-bu}
we see that the ratio of the zero mode
force to the mutual friction force is
$f_{\rm zm}/f_{\rm mf} \sim (\om_0\tau_c \pi^2\al^2)^{-1}$, 
assuming $\al_s\sim 1$. Using the estimates given above,
\beq
\frac{f_{\rm zm}}{f_{\rm mf}} \sim
0.003 \,
\Bigl(\frac{\mu_q}{400~\MeV}\Bigr)^{\!1/3}
\Bigl(\frac{50~\MeV}{\De}\Bigr)^{\!2}
\Bigl(\frac{T}{0.01~\MeV}\Bigr)^{\!5/3} \ .
\label{zm-ratio}
\eeq
We conclude that the zero mode force is likely to be
negligible relative to mutual friction.

\subsection{Timescale for expulsion of flux}

We can now estimate the time scale for the expulsion
of the $X$ magnetic field flux tubes from the 2SC core.
As we noted above, there will be an outward force on the flux
tubes at the point where they reach the nuclear mantle. The maximum
force per unit length is given by \eqn{fbmax}, in which the
energy costs of the magnetic field in the nuclear mantle have been
neglected. 
The rate of outward movement of the flux tubes is 
given by balancing that force against frictional or pinning
forces. There may be such forces arising from the nuclear
matter, but we ignore them and only include
the Aharonov-Bohm (mutual friction and Iordanskii)
and Magnus-Lorentz forces in the quark matter.
Using \eqn{eq:dynamics}, \eqn{eta-defn}, \eqn{Iordanskii},
\eqn{ML-final}, we can
see that the steady-state value of the vortex velocity $\vecv_L$
is given by the force balance equation,
\beq
\rho_{\rm ML}(\vecv_S-\vecv_L)\times \hat n
+D'(\vecv_L-\vecv_N)\times \hat n
+\eta (\vecv_L-\vecv_N)
+\vecf_{\rm bf}(r)=0 \ ,
\eeq
where $\vecf_{\rm bf}(r)$ is given by \eqn{fbmax}.
We work in the reference frame that is uniformly rotating with 
the normal component (blue quarks and electrons) and we neglect possible
small differential rotation between the superfluid and the normal
fluid, so $\vecv_N=\vecv_S=0$ in this frame.
The Iordanskii and Magnus-Lorentz forces then
add to give a single transverse force
\beq
-\rho\vecv_L\times \hat n +\eta \vecv_L + \vecf_{\rm bf}(r)=0 \ ,
\label{balance}
\eeq
where $\rho=\rho_{\rm ML}-D'$. From \eqn{ML-final} and Sec.~\ref{sec:Iordanskii}
we see that $\rho_{\rm ML}\sim\mu_q^3$ and $D'\sim \al\mu_q^3$, so
we can neglect the Iordanskii force and assume $\rho\approx \rho_{\rm ML}$.

We take the flux tube to lie in the $z$ direction,
and we calculate its position in the $x,y$ plane using polar co-ordinates
$(r,\th)$. We want to find $\dot r$, the rate at 
which the flux tube moves outward. Solving \eqn{balance} for the
steady-state velocities $\dot r$ and $\dot\th$, we find
\beq
\ba{rcl}
\dot r &=&\dsp \frac{\eta}{\eta^2+\rho^2} f_r(r) \ , \\[2ex]
r \dot\th &=&\dsp \frac{\rho}{\eta^2+\rho^2} f_r(r) \ ,
\ea
\label{fluxtube-EoM}
\eeq
where $f_r$ is the radial component of the boundary force.
We note in passing that $\dot r$ shows a {\em non-monotonic}
dependence on the friction coefficient $\eta$. As $\eta$ tends
to zero one might expect the expulsion time to also tend to zero,
and in the absence the Magnus-Lorentz force ($\rho=0$) this would
indeed be the case. However, in the presence of a non-zero
Magnus-Lorentz force, the 
flux tube moves in an orbit around the center of the star,
with the radially outward boundary force balanced by the resultant
radially inward Magnus-Lorentz force.

If the flux tube starts at radius $r_0$ at time $t=0$ and leaves
the core ($r$ reaches $R$) at time $t=t_1$, then by solving
\eqn{fluxtube-EoM} we find
\beq
\ba{rcl}
t_1 &=&\dsp \tau \biggl[
  2\ln\Bigl(\frac{R}{r_0}\Bigr) + 1-\frac{r_0^2}{R^2}
\biggr] \ , \\[3ex]
\tau &=&\dsp \frac{R^2}{2\ep_X}\frac{\eta^2 + \rho^2}{\eta} \ .
\ea
\label{t1}
\eeq
The factor in square brackets of order 1 for initial radii $r_0$
not too close to 0 or $R$, so the flux expulsion time for a typical
flux tube is of order $\tau$.
From \eqn{eq:eta} and \eqn{ML-final}, $\eta\sim\al^2\mu_q^3$ and
$\rho\sim \mu_q^3$. So $\rho\gg\eta$, and 
using \eqn{X-tension}, \eqn{fbmax}, \eqn{ABfactor-bu} we find 
\beq
\tau \approx 
\frac{8 \al_s^2\mu_q R^2}{\pi \al^2 \ln\kaX} \ .
\label{tau}
\eeq
Taking $\alpha_s\approx 1$,
\beq
\tau \approx (10^{10}\,{\rm yr}) 
 \Bigl(\frac{\mu_q}{400\,\MeV}\Bigr)
 \Bigl(\frac{R}{1\,\km}\Bigr)^{\!2} \frac{1}{\ln\kaX} \ .
\label{tau-approx}
\eeq
The timescale for $X$ flux tubes to be expelled from the 2SC core
is therefore in the range of $10^{10}$ years.

\section{Conclusions}
\label{sec:conclusions}

Quark matter in the 2SC (or CFL) color-superconducting phase 
is a superconductor
with respect to a broken ``$X$'' generator that is mostly color
with a small admixture of electromagnetism.
We have confirmed previous calculations \cite{Iida:2002ev}
showing that quark matter
in the 2SC phase will be a type-II $X$-superconductor if the
quark pairing gap is above a critical value which is well
within the expected range \eqn{kappa-2SC}.
Although the ambient magnetic field in the core of a neutron star
is below the lower critical field for the formation of
Abrikosov flux tubes containing $X$-magnetic flux, 
we argue that, when the quark matter 
cools into the 2SC phase, the process of domain formation and
amalgamation is likely to leave some of the $X$ flux trapped in the form
of flux tubes. The exact configuration and density of such tubes
depends on details of the dynamics of the phase transition,
but the density could be within an order of magnitude of the
density of conventional flux tubes in proton-superconducting
nuclear matter \eqn{eq:flux_number}.
Our calculations apply to 2SC quark matter in the temperature range
$T_{1SC}<T \ll T_{2SC}$ where $T_{2SC}$ is the critical temperature
for the formation of the 2SC condensate, expected to be
of order $10\,\MeV$ ($10^{11}\,\K$), and $T_{1SC}$ is the critical
temperature for self pairing of the blue quarks, which could be
as low as $1\,\eV$ ($10^4\,\K$). 

The 2SC phase contains three species of gapless fermions: two quarks
(``blue up'' and ``blue down'') and the electron. These are expected
to dominate its transport properties.
We do not discuss strange quarks,
but our analysis is also applicable to phases with 
strange quarks present, as long as their pairing pattern does not
break the $\Qt$ gauge symmetry. Muons may also be present, but, like
strange quarks, their higher mass gives them a lower Fermi momentum
so they make a subleading contribution to
the phenomena discussed here.
We have calculated the Aharonov-Bohm scattering cross-section
of gapless fermions with the $X$ flux tubes
\eqn{AB-scattering}, \eqn{ABfactor-bu},
and the associated collision (or relaxation) rate \eqn{tauinv-flux}.
A comparison with the collision time for Coulomb quark-quark scattering
indicates that only at very low temperatures 
($T\lesssim 10^5\,\K$ or $10\,\eV$) will the flux tubes
dominate over thermal relaxation via Coulomb scattering. However, we defer
a detailed calculation of the transport properties, including
Coulomb and $X$-boson-mediated interactions, to future work.

Because the ambient magnetic field in a neutron star is below
the lower critical field required to force $X$-flux tubes into
2SC quark matter, the trapped flux tubes will feel a boundary force
pulling them outwards. We calculated this force for the case where
the energy of the magnetic field outside the core can be
neglected relative to the energy of flux tube.
This force will be balanced
by the drag force (``mutual friction'') on the moving flux 
tube due to its Aharonov-Bohm interaction with the thermal population of
gapless quarks and electrons \eqn{eq:eta}, and also by 
the Magnus-Lorentz force \eqn{ML-final}. 
On this basis, we estimate that the timescale for the expulsion of
$X$ flux tubes from a 2SC core \eqn{tau-approx}, is of order
$10^{10}$  years. 

The work described here offers many directions for future 
development.\\
(1) To get a full picture of the transport properties of 2SC quark
matter one must calculate the relaxation rates associated with
processes that do not include flux tubes,
such as $\Qt$-Coulomb and $X$-boson-mediated
interactions between gapless fermions.\\
(2) We studied the regime where the cyclotron frequency is smaller than
the thermal relaxation time of the unpaired quarks 
(see Sec.~\ref{sec:Qt-field}). It would be valuable
to extend our analysis to higher magnetic fields and/or lower temperatures
where the cyclotron frequency cannot be neglected.\\
(3) It is important to resolve the disagreement in the literature
over whether the Magnus-Lorentz force on flux tubes is cancelled
by forces arising from the neutralizing background (see Sec.~\ref{sec:ML}).
This is necessary for understanding flux expulsion
from superconducting nuclear matter as
well as more exotic flux tubes such as the ones that we described here.\\
(4) We assumed that the $X$-flux tubes are stable, or at least
have a lifetime that is long enough for them to play a role in transport.
However there is no topological guarantee of their stability, and it
is necessary to perform a calculation of their energetics, 
analogous to that of \cite{James:1992wb}, and to investigate
bound states on the string, which if present can enhance their
stability \cite{Vachaspati:1992mk}.
\\
(5) We focussed on the 2SC phase, but other phases may support
flux tubes. The CFL phase, which is the ground state of 3-flavor
quark matter at asymptotically high densities, also has a gauge symmetry
breaking pattern which resolves an external magnetic field 
in to an unbroken $\Qt$ part, and a broken $X$ part which
could be carried in flux tubes \cite{Iida:2004if}. 
In this case also there is
no topological guarantee of stability, and an analysis of the
energetic stability is required.
The CFL phase also
features semi-superfluid vortices with non-zero 
magnetization~\cite{Iida:2002ev,Balachandran:2005ev,Eto:2009bh,Eto:2009wu,Sedrakian:2008ay}.
Since the CFL phase has no gapless charged excitations 
the associated phenomenology is likely to be quite different. 
In the CFL-K0 phase there are charged kaon modes that can have an energy
gap well below the pairing gap, so, if they have non-zero Aharonov-Bohm
$\tilde\beta$ factors, their scattering off flux tubes might be important.
\\
(6) We treated the thickness of the flux tubes as negligible, so
scattering off them is dominated by the Aharonov-Bohm effect.
In fact the thickness of the flux tube
is comparable to the inverse Fermi momentum
of the quarks (see \eqn{lambda-X})
and there will be finite-size corrections to our
results. Calculating them would require explicit construction of
the radial profile of the flux tube.\\
(7) Some quark matter phases break the $\Qt$ gauge symmetry. These
include the 2SC phase at $T<T_{1SC}$, and many other phases
such as the color-spin-locked phase \cite{Schafer:2000tw,Schmitt:2003xq}.
It is interesting to ask what happens to magnetic flux
in such cases: is the $\Qt$-superconductivity always type-I? (One
suspects it may be because the gaps are usually small.) 
Will the dynamics of the phase transition lead to
trapped normal regions, and what is the timescale for their expulsion
from the star? Could these phases retain $X$-flux tubes even after
$\Qt$ flux has been expelled? If $X$-flux tubes existed in a CFL core, for
example they might experience the same sort of entanglement with superfluid
vortices as is predicted in nuclear matter.\\
(8) Neutron stars probably have layers of different phases. For
a proper treatment of the dynamics of magnetic flux
one would have
to analyse how magnetic flux was connected between layers
and pinned within layers, and the consequent additional forces
on the color magnetic flux tube in a 2SC core.
For instance, in a conducting nuclear mantle there would be
eddy-current pinning of the magnetic flux; in a type-II superconducting
{\em and} superfluid mantle there would be entanglement of
nuclear Abrikosov flux tubes with  superfluid vortices; and so on.
There is also the possibility of different quark
matter phases, such as an inner CFL core, {\em inside} the 2SC
region. If it turned out that additional forces arising from
these other regions of the star acted so as
to allow expulsion of the flux tubes on a shorter timescale
then this
would have interesting astrophysical ramifications, such as
a change of the magnetic moment 
of the star over this period of time. If the core contained a phase
where $X$-flux tubes were entangled with superfluid vortices
(as mentioned for the CFL phase above) then the rotational dynamics 
could also be affected.
Observationally, this could provide 
a new mechanism for glitches in neutron stars, since vortex-interface 
pinning force, derived above, may prevent a continuous flow of rotational 
vortices in the superfluid phases, in a manner analogous to 
vortex pinning in the crust \cite{Anderson:1975zze} and the hadronic 
core-solid crust interface~\cite{Sedrakian:1998ki}. 
Other dynamical manifestations, such as, for example, the
recently studied shear modes~\cite{Noronha:2007qf,Shahabasyan:2009zz} 
in the superfluid core and the post-jump relaxations 
(see Ref.~\cite{Sedrakian:2006xm} and references therein) will be affected as 
well.

\section*{Acknowledgements}
We thank  Xu-Guang Huang, Kazunori Itakura, Naoki Itoh,
Peter Jones, Muneto Nitta, Dirk Rischke, Karen Shahabasyan
for their comments.
This research was
supported in part by the Offices of Nuclear Physics and High
Energy Physics of the U.S.~Department of Energy under contracts
\#DE-FG02-91ER40628,  
\#DE-FG02-05ER41375, 
and the Deutsche Forschungsgemeinschaft (Grant SE 1836/1-1).

\appendix  

\section{Relating the scattering amplitude to Aharonov-Bohm cross-section}
\label{app:corss_section}

The differential scattering probability is given by~\cite{BLP:1982} 
\be\label{eq:w1}
dW = 2\pi\delta(\ep'-\ep) 2\pi\delta(p_z-p_z')
\vert M_{fi}\vert^2 \frac{1}{2\ep V}
\prod_a \frac{d^3p'_a}{(2\pi)^32\ep_a'}.
\ee
It is assumed that a particle scatters off a heavy center; the 
momentum conserving delta function reflects the fact that there
is no scattering  along the vortex (\ie, in the $z$
direction). The 
quantities referring to the final states are primed, those to 
the initial state are unprimed. The $a$ products is over all the 
final state particle phase space. The differential scattering 
cross section is 
\be\label{sig1} 
d\sigma = \frac{d W}{j}, \quad j = \frac{V}{v} 
= \frac{\ep V}{\vert\vecp\vert},
\ee
here $j$ is the current, $v$ the velocity, and $V$ the volume. 
Substituting (\ref{sig1}) in Eq.~(\ref{eq:w1}) we obtain for
the scattering cross-section 
\be 
d\sigma = j^{-1}2\pi\delta(\ep'-\ep) 2\pi\delta(p_z-p_z')
\vert M_{fi}\vert^2 \frac{1}{2\ep V}
\prod_a \frac{d^3p'_a}{(2\pi)^32\ep_a'}.
\ee
The differential scattering cross section is obtained
on writing $d^3p' = d\pperp' \pperp'd\phi' dz'$  (we 
restrict in the following the $a$ product to one particle, 
since we consider elastic scattering and there is the 
same blue quark or electron in the final state). Thus,
\be 
 \frac{d\sigma}{d\phi'}= j^{-1}2\pi\delta(\ep'-\ep) 2\pi\delta(p_z-p_z')
\vert M_{fi}\vert^2 \frac{1}{2\ep V} 
\frac{d\pperp' \pperp' dp_z'}{(2\pi)^32\ep'}.
\nonumber\\
\ee
The final state energy is $\ep'= \sqrt{\pperp'^2+p_z'^2}$, 
therefore $\ep'd\ep' = \pperp' d\pperp'$.
After integrating by means of delta-functions we obtain
\bea \label{cross_section}
\frac{d\sigma}{d\phi'} &=& j^{-1}
\vert M_{fi}\vert^2 \frac{1}{8\pi\ep V} .
\eea
The current density is given by $j = v_{\perp} /L^2$, where $v_{\perp}= 
\vert\pperp\vert/\ep$~\cite{BLP:1982}. Comparing 
(\ref{cross_section}) with Eq.~(\ref{AB-scattering}) we obtain
\be\label{M_squared}
\vert M_{fi}\vert^2 = \frac{4 L\sin^2(\pi\tilde\beta)}{\sin^2(\phi/2)}.
\ee
Finally, substituting this result in Eq.~(\ref{eq:w1}) we arrive 
at Eq.~(\ref{eq:diff_probability}) of the main text.

\section{Phase space integrals}
\label{app:phase_space}
Upon substituting the transition probability in the rate Eq.~(\ref{eq:s1})
and introducing momentum transfer $\veck = \vecp'-\vecp$ we find
\bea\label{eq:s2}
s_0 &=&\frac{4\pi}{\hbar} 
\frac{n_v}{{\cal D}} \sin^2(\pi\tilde\beta) \int\!\!\frac{d^3k}{(2\pi)^2} 
\left[\veck\cdot \vecv\right]^2\delta(k_z)~ I_p
\eea
where we replaced $N_vL/V=n_v$, which is the density of flux tubes
per unit area. The integral $I_p$ is defined as 
\bea
I_p &=&\int\frac{d^3p}{(2\pi)^3}f_0(\vecp)\left[1-f_0(\vecp)\right] 
\delta(\ep(\vecp)-\ep(\veck_{\perp}-\vecp))
\frac{1}{\sin^2(\phi/2)} \frac{1}{4\ep(\vecp)^2}.
\eea
The form of the scattering probability suggest that 
the phase space integral over $d^3p$ is convenient to carry out in 
the cylindrical coordinates by writing $d^3p = p_{\perp}dp_{\perp} 
d \phi d p_z$ and 
\bea
I_p &=&
\frac{1}{4(2\pi)^3}\int dp_{\perp}p_{\perp} 
d \phi \frac{1}{\sin^2(\phi/2)}
\int d p_z f_0(\vecp)\left[1-f_0(\vecp)\right]
\frac{1}{\ep(\vecp)^2}
\delta(\ep(\vecp)-\ep(\vecp-\veck_{\perp})).\nonumber\\
\eea
To do the  inner integral note that in the low-temperature 
limit 
\be 
 f_0(\vecp)\left[1-f_0(\vecp)\right] 
= T\frac{df_0(\vecp)}{d\ep(\vecp)}
\simeq T\frac{p_F}{\sqrt{p_F^2 -p_{\perp}^2}}~\delta(p_z-p_z^0) \ ,
\ee
where in the last step we used cylindrical polar coordinates with
$p_z^0 = \sqrt{\ep_F^2 -p_{\perp}^2}$.
Carrying out the $p_z$-integration we obtain
\bea\label{eq:ip1}
I_p=-\frac{T}{4(2\pi)^3}
\int\!\!\frac{dp_{\perp} p_{\perp}}{p_F\sqrt{p_F^2 -p_{\perp}^2}}
\theta(\sqrt{p_F^2 -p_{\perp}^2}) I_{\phi},
\eea
where
\bea
I_{\phi}=
\int\!\!\left[\frac{ d \phi}{ \sin^2(\phi/2)}\right]
\delta(\ep_F- \sqrt{p_F^2+ k_{\perp}^2
-2\vecp_{\perp} \cdot \veck_{\perp}}).
\eea
We next specify the geometry of the scattering, by 
assuming that the vortex is along $z$-axis, vector $p$ 
is directed along the $x$ axis, and the scattering is 
in the $x$-$y$ plane. If we denote the angle formed by the 
vectors $\vecp_{\perp}$ and $\veck_{\perp}$ by $\chi$ then, 
$\cos \chi = \sin \phi/2$. We next note that identically
$d\phi = -2d\sin(\phi/2)/\cos(\phi/2)$ and define
\be
 \sin \frac{\phi_0}{2} = \frac{\kperp}{2\pperp}.
\ee
The integral becomes
\bea 
I_{\phi} &=& -2\frac{p_F}{p_{\perp}k_{\perp}}
\int_{-1}^{1} \frac{d\sin(\phi/2)}{\cos(\phi/2)} 
\left[\frac{1}{\sin^2(\phi/2)}\right]
\delta(\sin \phi/2 - \sin \phi_0/2) \nonumber\\
&=& -\frac{16p_F\pperp^2}{k_{\perp}^3}
\frac{1}{\sqrt{4\pperp^2-\kperp^2}} 
\theta\left(1-\frac{\kperp}{2\pperp}\right).
\eea
The integration limits were  chosen for convenience $[-1;1]$; 
the integral is then multiplied by a factor 2 to account for 
full 360 degree angle range. Inserting these results in 
Eq.~(\ref{eq:ip1}) we obtain
\bea
I_p&=&\frac{2T}{(2\pi)^3k_{\perp}^3}
\int\!\!\frac{dp_{\perp} p_{\perp}^3}{\sqrt{p_F^2 -p_{\perp}^2}}
\frac{1}{\sqrt{\pperp^2-\frac{\kperp^2}{4}}} 
\theta\left(1-\frac{\kperp}{2\pperp}\right)
\theta(\sqrt{p_F^2 -p_{\perp}^2}).
\eea
The integration is carried out using 
\be 
\int_a^bdx \frac{x^3}{\sqrt{b^2-x^2}\sqrt{x^2-a^2}} = 
\frac{\pi}{4}\left(b^2+a^2\right),
\ee
to obtain the final expression for the momentum integral
\bea
I_p&=&\frac{T}{16\pi^2 k_{\perp}^3}\theta\left(1-\frac{\kperp}{2p_F}\right)
\left(p_F^2+\frac{\kperp^2}{4}\right).
\eea
Next we substitute this result in Eq.~(\ref{eq:s2}) and obtain
\bea\label{eq:s3}
s_0 &=&\frac{n_v v^2T}{4\pi\hbar{\cal D}} 
 \sin^2(\pi\tilde\beta) \int_0^{2p_F}\frac{d\kperp }
{(2\pi)^2}  
\left(p_F^2+\frac{\kperp^2}{4}\right)
\int_0^{2\pi} d\phi (\cos\phi)^2.
\eea
The angular integral is equal $\pi$, 
the remaining integral is $(8/3)p_F^3$, and
we arrive at Eq.~(\ref{eq:s5}) of the main text.

\renewcommand{\href}[2]{#2}
\newcommand{\apjl}{Astrophys. J. Lett.}
\newcommand{\mnras}{Mon. Not. R. Astron. Soc.}

\bibliographystyle{JHEP_MGA}
\bibliography{2SC_flux_tube} 

\end{document}